\documentclass[%
 reprint,
 amsmath,amssymb,
 aps,
]{revtex4-2}

\usepackage{graphicx}% Include figure files
\usepackage{dcolumn}% Align table columns on decimal point
\usepackage{bm}% bold math
\usepackage{bbm}
\usepackage{hyperref}% add hypertext capabilities
\usepackage{parskip}

\usepackage{tabularx}

\begin{document}

\preprint{APS/123-QED}

\title{Divine Social Networking in the Age of Lost Omens}% Force line breaks with \\

\author{W. Brian Lane$^{1,2}$}
%  \altaffiliation{Department of Physics, University of North Florida.}%Lines break automatically or can be forced with \\
 \email{Brian.Lane@unf.edu}
\affiliation{%
 $^{1}$Department of Physics, University of North Florida\\
 $^{2}$Northeast Florida Center for STEM Education, University of North Florida\\
 1 UNF Drive, Jacksonville, FL, 32224% \textbackslash\textbackslash
}%

\date{\today}% It is always \today, today,

\begin{abstract}

The last two years have seen significant changes in the divine pantheon of the Lost Omens campaign setting of the Pathfinder Tabletop Roleplaying Game. First, the \textit{Pathfinder Remaster}, necessitated by the Open Game License debacle, prompted the removal of alignment and an enrichment of divine identities and relationships. Second, the \textit{War of Immortals}, kicked off by the death of one of the core 20 deities, shook up the membership and relationships within the setting's primary pantheon. These two changes prompted the reprinting of deity information in \textit{Pathfinder: Lost Omens Divine Mysteries}, which updates and replaces the pre-Remaster \textit{Pathfinder: Lost Omens Gods \& Magic}. Notably, \textit{Divine Mysteries} features double the page count profling the core 20 deities. In this paper, we use \textit{social network analysis} to examine the impact of these changes (Remaster, War of Immortals, and page count) on the relationships among the core 20 deities. In this analysis, each deity features as a node, connected by edges that represent the number of times each pair of deities is mentioned in each other's profiles. The results reveal a much richer, more connected divine network in \textit{Divine Mysteries} than in \textit{Gods \& Magic}. We conclude by discussing implications for the Lost Omens campaign setting and areas of future development.

% \begin{description}
% Secondary publications and information retrieval purposes.
% You may use the \texttt{description} environment to structure your abstract;
% use the optional argument of the \verb+\item+ command to give the category of each item. 
% \end{description}
\end{abstract}

\maketitle

%\tableofcontents

\section{\label{sec:intro}Deities in the Lost Omens Campaign Setting}

Deities are one of the most celebrated aspects of lore in the Lost Omens campaign setting for the Pathfinder Tabletop Roleplaying Game. With deities representing various cultures, ideologies, LGBTQ+ identities, hobbies, and professions, it is never difficult to find a deity one's player character can pledge to. During the past two years, the Lost Omens pantheon has seen significant shifts.

First, in late 2022 and early 2023, leaks revealed proposed changes to the Open Game License (OGL) that Pathfinder was published under at the time \cite{Hoffer2022Dungeons}. These changes threatened to destabilize the Pathfinder brand, prompting publisher Paizo to remaster Pathfinder under a new Open RPG Creative License \cite{Codega2023Paizo}. The Remaster produced an updated edition that left behind certain elements that could be uniquely claimed under the OGL. One such element that greatly affected the deities of the Lost Omens campaign setting was \textit{alignment}. Under alignment, each deity is described by two traits: one referring to moral quality (good, neutral, or evil) and a second referring to orderliness (lawful, neutral, or chaotic). These two traits operate independently, producing 9 possible descriptors for a deity's disposition. When originally developing 20 deities for the Lost Omens core pantheon, Paizo took care to make at least 2 deities available with each trait combination, ensuring diverse representation across a discretized moral system. Without alignment, the Lost Omens deities require more nuanced discussion.

Second, in 2023, Paizo announced that one deity from among the core 20 would die in an upcoming major setting event. The doomed deity was revealed in 2024 to be Gorum, the setting's god of battle, strength, and weapons, and his death was formalized in various adventures and the \textit{War of Immortals} source book. The events surrounding Gorum's death resulted in other changes to the pantheon, most notably the promotion of Arazni to a place of prominence among the core 20 deities.

As a result of both these changes, Paizo published in 2024 \textit{Pathfinder: Lost Omens Divine Mysteries} (LODM \cite{LODM}), an update to the pre-Remaster \textit{Pathfinder: Lost Omens Gods \& Magic} (LOGM \cite{LOGM}) that offered detailed information about the core 20 deities. Among other information, LODM and LOGM detail relationships between these deities, with two notable differences in structure. First,  LODM devotes twice as much page count (4 pages as opposed to 2) as LOGM. This expansion permits more nuanced discussions of divine relationships. Second, LOGM summarized each deity's allies, enemies, and relationships in a header with ``Allies,'' ``Enemies,'' and ``Relationships,'' while LODM integrated this information within the deity's descriptive text.

In this paper, we quantitatively analyze this descriptive text to answer the following research Questions: 

\textbf{RQ1}: How did the network of divine relationships change between LOGM and LODM? 

\textbf{RQ2}: How can we explore specific influences of the Pathfinder Remaster, Gorum's death, and increased page count on changes in the divine network?

We answer these questions using \textit{social network analysis} (SNA) as described in Section \ref{sec:methodology}. We present the results of this SNA in Section \ref{sec:results} and discuss findings and implications for the future of the Lost Omens campaign setting in Section \ref{sec:discussion}.

\section{\label{sec:methodology}Methodology}

\begin{figure}
    \centering
    \includegraphics[width=0.5\linewidth]{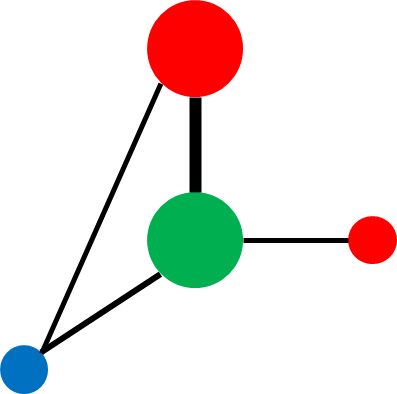}
    \caption{A sample network, consisting of 4 nodes (circles) and 4 edges (lines).}
    \label{fig:sample-network}
\end{figure}

SNA provides methods of holistically examining relationships among a social network of individuals. In the language of graph theory, each individual is represented as a \textit{node} and a documented connection between two individuals is represented as an \textit{edge} connecting those nodes. For example, Figure \ref{fig:sample-network} shows a simple network of 4 nodes (circles) and 4 edges (lines). Information about the nodes and edges can be encoded visually such as by the color, shape, and size of the node and the thicnkess of the edges. In fact, LODM features a network diagram in its Divine Connections Map, with edge patterns indicating relationships of allies, enemies, lovers, and others.

Once data is collected about nodes and edges, SNA provides quantitative tools to examine features of a network and differences between two networks. In this section, we outline how we collected social network data about the Lost Omens core 20 deities from LOGM and LODM and the SNA techniques we can use to compare the networks generated from these data sources.

\subsection{Data Collection}

In the networks discussed here, nodes represent the Lost Omens core 20 deities and edges represent their relationships. To establish relationships between deities, we counted the number of times $n_{ij}$ that deity $i$'s profile mentions deity $j$. For symmetry, we define the elements of an \textit{adjacency matrix} as the sum

\begin{equation}
    d_{ij} = n_{ij} + n_{ji}.
\end{equation}

In our SNA, we use $d_{ij}^{\alpha}$ as the \textit{weight} of the edge between deities $i$ and $j$ in network $\alpha$. In our case, $\alpha = \textrm{LOGM},\textrm{LODM}$. We recorded these $20 \times 19 / 2 = 190$ values in a spreadsheet for both LOGM ($d_{ij}^{\textrm{LOGM}}$) and LODM ($d_{ij}^{\textrm{LODM}}$), with the definition $d_{ii}^{\alpha} = 0$ (no edge between a deity and themself). 

\subsection{\label{subsec:sna}Social Network Analysis Metrics}

SNA offers several metrics we can use to evaluate whole-network similarities and differences between $d_{ij}^{\textrm{LOGM}}$ and $d_{ij}^{\textrm{LODM}}$:

\textbf{Node Degree:} The \textit{degree} $D_{i}^{\alpha}$ of node $i$ in network $\alpha$ is defined as the number of edges that connect $i$ to another node. Defining $\mathbbm{1}(n>0)$ as $1$ when $n>0$ and $0$ otherwise, 

\begin{equation}
    D_{i}^{\alpha} = \sum_{j \neq i} \mathbbm{1}\left(d_{ij}^{\alpha}\right).
\end{equation}

Degree is a measure of how many connections a node has within the network.

\textbf{Node Strength}: The \textit{strength} $S_{i}^{\alpha}$ of node $i$ in network $\alpha$ is defined as the total weight of all the edges that connect $i$ to another node:

\begin{equation}
    S_{i}^{\alpha} = \sum_{j \neq i} d_{ij}^{\alpha}.
\end{equation}

Strength is a measure of how robust a node's connections are to the rest of the network.

\textbf{Node Betweenness}: A node's \textit{betweenness} $B_{i}^{\alpha}$, expressed as a number between 0 and 1, is a fractional count of how many times a node is found along the shortest path linking two other nodes in network $\alpha$:

\begin{equation}
    B_{i}^{\alpha} = \frac{\sum_{j \neq k \neq i} s_{jk}^{\alpha i}}{\sum_{j \neq k}^{\alpha} s_{ij}},
\end{equation}

where $s_{jk}^{\alpha}$ is the number of shortest paths between nodes $j$ and $k$ in network $\alpha$ and $s_{jk}^{\alpha i}$ is the number of shortest paths between nodes $j$ and $k$ that pass through node $i$. In a network with weighted edges like we consider in this paper, path lengths are computed using the inverse of an edge weight as distance.

\textbf{Node Closeness}: A node's \textit{closeness} $C_{i}^{\alpha}$ is a measure of the node's proximity to other nodes in network $\alpha$, measured as 

\begin{equation}
    C_{i}^{\alpha} = \frac{N^{\alpha}-1}{\sum_{j} l_{ij}^{\alpha}},
\end{equation}
where $l_{ij}^{\alpha}$ is the distance between nodes $i$ and $j$ in network $\alpha$, based on inverse edge weight as discussed above.

\textbf{Density}: The \textit{density} $\rho^{\alpha}$ of a network $\alpha$ is defined as the number of edges that exist within the network as a fraction of the number of edges that could exist:

\begin{equation}
    \rho^{\alpha} = \frac{\sum_{ij} \mathbbm{1} \left( d^{\alpha}_{ij} > 0 \right)}{N^{\alpha}(N^{\alpha}-1)/2}.
\end{equation}

Density is a measure of how well connected an entire network is.

\textbf{Node Degree Cosine}: The \textit{node degree cosine} $\textrm{NDC}^{\alpha,\beta}$ between two networks $\alpha$ and $\beta$ is defined as

\begin{equation}
    \textrm{NDC}^{\alpha,\beta} = \frac{\sum_{i}D_{i}^{\alpha}D_{i}^{\beta}}{\sqrt{(\sum_{i}(D_{i}^{\alpha})^{2})(\sum_{i}(D_{i}^{\beta})^{2})}}.
\end{equation}

This quantity is named ``cosine'' because of its structural similarity to an inner product divided by a product of magnitudes. NDC measures how similar the set of node degrees are. Two identical networks will have an NDC of 1. Two networks with no edges in common will have an NDC of 0.

\textbf{Node Strength Cosine}: The \textit{node strength cosine} $\textrm{NSC}^{\alpha,\beta}$ between two networks $\alpha$ and $\beta$ is defined as

\begin{equation}
    \textrm{NSC}^{\alpha,\beta} = \frac{\sum_{i}S_{i}^{\alpha}S_{i}^{\beta}}{\sqrt{(\sum_{i}(S_{i}^{\alpha})^{2}) (\sum_{i}(S_{i}^{\beta})^{2})}}.
\end{equation}

Similar to NDC, NSC measures how similar the set of node strengths are. Two identical networks will have an NSC of 1. Two networks with no edges in common will have an NSC of 0.

\textbf{Edge Existence Jaccard}: The \textit{edge existence Jaccard} $\textrm{EEJ}^{\alpha,\beta}$ is a count of the number of edges two networks have in common as a fraction of the total number of edges in their union:

\begin{equation}
    \textrm{EEJ}^{\alpha,\beta} = \frac{\left| E^{\alpha} \cap E^{\beta} \right|}{\left| E^{\alpha} \cup E^{\beta} \right|}
\end{equation}

where $E^{\alpha}$ is the set of edges in network $\alpha$. Two identical networks will have an EEJ of 1. Two networks with no edges in common will have an EEJ of 0.

\subsection{Social Network Analysis Clustering}

A \textit{clustering algorithm} provides a means of identifying subsets of nodes within a network that are more tightly connected to each other than they are to nodes outside the subset. There are many clustering algorithms available. We choose the \textit{Louvain algorithm} \cite{Louvain} since it accounts for the weight of connections between nodes, which is an important feature in the LOGM and LODM networks.

\section{\label{sec:results}Results}

Here we present the results of our network comparison based on the $d_{ij}^{\alpha}$ values identified from LOGM and LODM. We begin by highlighting features of the network diagrams. Then, we discuss network comparison metrics between LOGM and LODM. Finally, we analyze network clusters and two sub-networks that involve only the 19 deities that these two networks have in common.

\subsection{Social Network Diagrams}

\begin{figure*}
    \centering
    \includegraphics[width=0.80\textwidth]{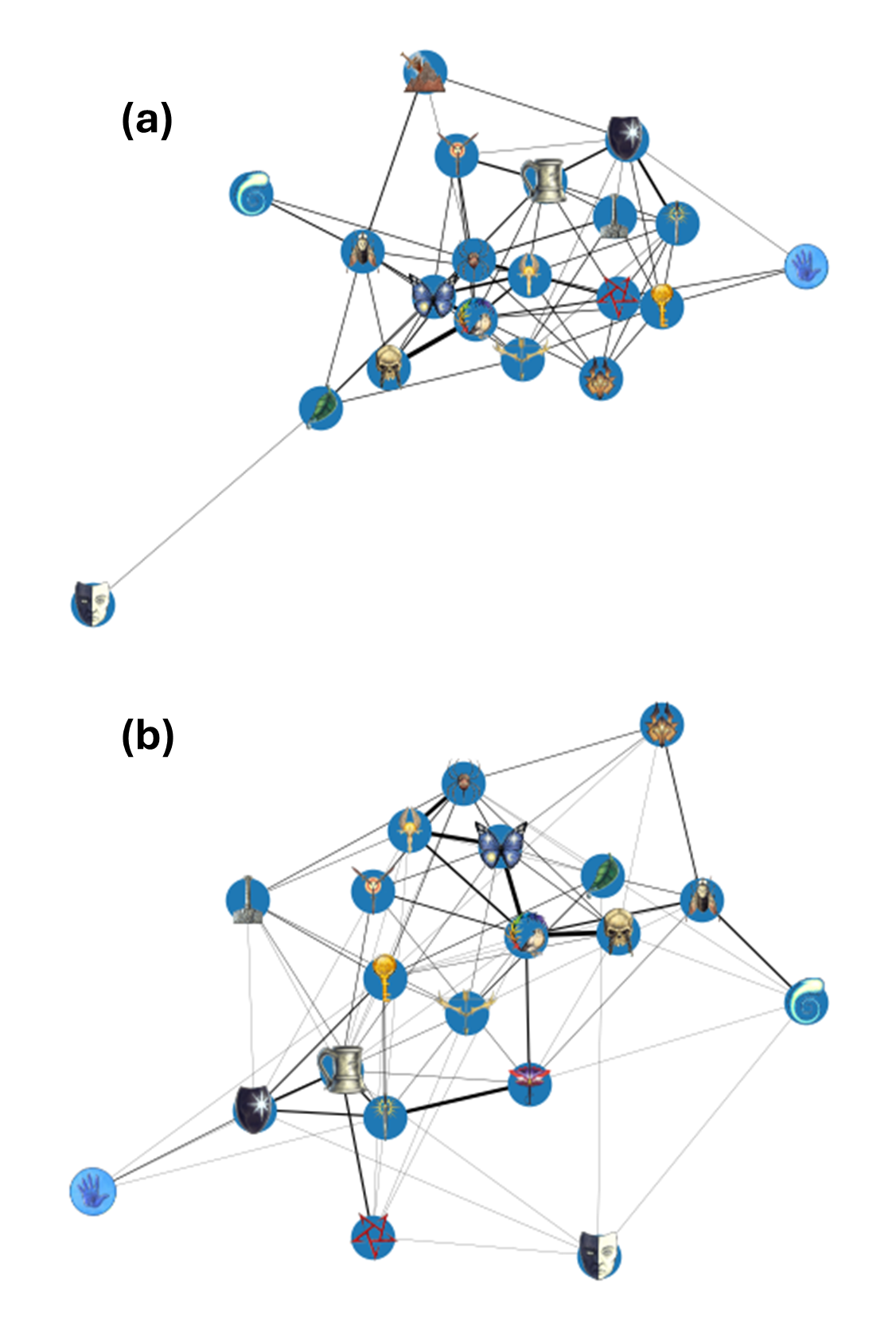}
    \caption{Social network diagrams created based on deity profiles in (a) \textit{Pathfinder: Lost Omens Gods \& Magic} and (b) \textit{Pathfinder: Lost Omens Divine Mysteries}. The thickness of each edge between deity icons represents the number of times each deity's profile mentions the other.}
    \label{fig:network-diagrams}
\end{figure*}

The social network diagrams resulting from LOGM and LODM are shown in Figure \ref{fig:network-diagrams}(a) and (b), respectively. Here we discuss a few qualitative observations.

\subsubsection{Lost Omens Gods \& Magic Network Observations}

\textbf{Rovagug's Centrality}. Given Rovagug's importance in the origin story of the Lost Omens campaign setting and the existential threat he poses to the world, it is unsurprising to see Rovagug so well connected within the network diagram. Most notably, he is well connected with the older deities who allied to imprison him in the Dead Vault.

\textbf{Relative Isolation for Gorum, Irori, Nethys, and Pharasma.} In contrast, Gorum, Irori, Nethys, and Pharasma are less connected, each with three or fewer edges. It therefore seems unsurprising that Gorum would be selected for removal from the setting.

\textbf{The Radiant Prism}. One of the most fan-celebrated aspects of the Lost Omens setting, the Radiant Prism (romantic relationship among Desna, Sarenrae, and Shelyn) is manifested as three thick edges between these goddesses.

\textbf{Sibling Connections}. Another important relationship among the gods is the siblinghood between Shelyn and Zon-Kuthon, which manfiests as a thick edge between the two deities. Note that our methodology did not include references to Zon-Kuthon's previous name, Dou-Bral. Including these references would increase the weight of this edge.

\textbf{The Ascended}. Three of the core 20 deities---Cayden Cailean, Iomedae, and Norgorber---were raised from mortal humanity. This disparate group is collectively referred to as the Ascended. This group is somewhat recognizable in the upper-right corner of the diagram.

\subsubsection{Lost Omens Divine Mysteries Network Observations}

\textbf{Increased connections}. Unsurprisingly, we see much richer connections among the deities in LODM, which devoted twice as much page count to each deity. This additional space permitted discussion of more interactions or relationships between deities. In particular, Irori, Nethys, and Pharasma now feature more edges.

\textbf{Rovagug's Centrality}. Rovagug again enjoys strong connections within the LODM network, especially with the older deities.

\textbf{The Radiant Prism}. The Radiant Prism is still clearly displayed.

\textbf{Sibling Connections}. Shelyn and Zon-kuthon's familial connection is still clearly displayed.

\textbf{The Ascended}. The Ascended receive greater comparative discussion between each other in LODM, resulting a more prominently displayed triangle of edges.

\textbf{Arazni's Limited Connections}. LODM emphasizes that Arazni is somewhat reluctant to claim godhood, with limited relationships with other deities. Her strong connections with Iomedae and Shelyn indicates she is forming supportive connections in her newfound prominence.

\subsection{Network Comparison Metrics}

The visual layout of the network diagrams in Figure \ref{fig:network-diagrams} indicates notable differences in their structure, which we quantify here with the network comparison metrics outlined in Section \ref{subsec:sna}. The LOGM network (Figure \ref{fig:network-diagrams}(a)) has a density of $\rho^{\textrm{LOGM}} = 0.337$ (i.e., it has one-third of the possible edges it could have) and the LODM network (Figure \ref{fig:network-diagrams}(b)) has a higher density of $\rho^{\textrm{LODM}} = 0.416$.

Each deity's connectedness to the rest of the network is quantified by the centrality measures of bewteenness and closeness, presented in Table \ref{tab:centralities} for each network LOGM and LODM. In the LOGM network, most closeness values are between about 0.010 and 0.018. Nethys' minimal value of 0.006 is the only notable outlier and is reflected in his node's peripheral position in Figure \ref{fig:network-diagrams}(a). This near-uniformity indicates a similar average distance between one deity and the others in the network.

The deities' betweenness values in LOGM show greater disparity, with Shelyn and Rovagug holding the highest values of betweenness. This prominence indicates that the shortest path between any two deities is most likely to pass through Shelyn (an ancient goddess, member of the Radiant Prism, and sister of Zon-Kuthon) or Rovagug (the primordial existential threat that the older deities allied against). Notably, Irori, Lamashtu, Nethys, Pharasma, and Zon-Kuthon all have a betweenness of exactly zero, indicating that no shortest paths between two deities pass through any of them.

\begin{table*}
    \centering
    \begin{tabular}{c|cc|cc|cc|cc}
              & \multicolumn{2}{c|}{LOGM}                   & \multicolumn{2}{c|}{LODM} & \multicolumn{2}{c|}{LOGM sans Gorum} & \multicolumn{2}{c}{LODM sans Arazni} \\
        Deity & Closeness & Betweenness & Closeness & Betweenness & Closeness & Betweenness & Closeness & Betweenness  \\ \hline
Abadar & 0.013 & 0.031 & 0.021 & 0.099 & 0.014 & 0.035 & 0.02 & 0.157\\
Arazni & -- & -- & 0.025 & 0.246 & -- & -- & -- & --\\
Asmodeus & 0.015 & 0.076 & 0.017 & 0.0 & 0.015 & 0.078 & 0.015 & 0.0\\
Calistria & 0.013 & 0.018 & 0.02 & 0.0 & 0.013 & 0.026 & 0.02 & 0.15\\
Cayden Cailean & 0.014 & 0.053 & 0.022 & 0.123 & 0.014 & 0.065 & 0.019 & 0.225\\
Desna & 0.017 & 0.123 & 0.026 & 0.082 & 0.017 & 0.144 & 0.022 & 0.098\\
Erastil & 0.012 & 0.006 & 0.014 & 0.0 & 0.012 & 0.007 & 0.014 & 0.0\\
Gorum & 0.011 & 0.012 & -- & -- & -- & -- & -- & --\\
Gozreh & 0.012 & 0.105 & 0.016 & 0.006 & 0.012 & 0.111 & 0.016 & 0.007\\
Iomedae & 0.013 & 0.035 & 0.025 & 0.251 & 0.012 & 0.026 & 0.017 & 0.013\\
Irori & 0.009 & 0.0 & 0.013 & 0.0 & 0.01 & 0.0 & 0.012 & 0.0\\
Lamashtu & 0.012 & 0.0 & 0.017 & 0.0 & 0.012 & 0.0 & 0.015 & 0.0\\
Nethys & 0.006 & 0.0 & 0.008 & 0.0 & 0.006 & 0.0 & 0.008 & 0.0\\
Norgorber & 0.013 & 0.053 & 0.023 & 0.158 & 0.012 & 0.02 & 0.018 & 0.167\\
Pharasma & 0.011 & 0.0 & 0.018 & 0.041 & 0.011 & 0.0 & 0.015 & 0.046\\
Rovagug & 0.017 & 0.208 & 0.022 & 0.079 & 0.018 & 0.239 & 0.021 & 0.147\\
Sarenrae & 0.016 & 0.058 & 0.023 & 0.105 & 0.017 & 0.065 & 0.021 & 0.111\\
Shelyn & 0.018 & 0.228 & 0.027 & 0.453 & 0.018 & 0.229 & 0.022 & 0.33\\
Torag & 0.013 & 0.005 & 0.016 & 0.0 & 0.014 & 0.005 & 0.016 & 0.0\\
Urgathoa & 0.014 & 0.105 & 0.022 & 0.205 & 0.013 & 0.039 & 0.019 & 0.196\\
Zon-Kuthon & 0.015 & 0.0 & 0.024 & 0.0 & 0.015 & 0.0 & 0.02 & 0.0\\

        \end{tabular}
    \caption{Centrality measures (closeness and betweenness) for each of the 21 deities in LOGM and LODM.}
    \label{tab:centralities}
\end{table*}

Within the LODM network, each deity's closeness value is higher than in LOGM. With more connections described between deities, the path between any two deities becomes shorter. The largest value is Shelyn's 0.027, followed by Desna's 0.026 and then Iomedae and newcomer Arazni at 0.025.

Betweenness values, however, show differing shifts from LOGM to LODM. Iomedae's betweenness increases by a factor of 7, and Abadar's and Norgorber's betweenness increase by a factor of 3. Meanwhile, Asmodeus', Calistria's, Erastil's, and Torag's betweenness decrease to zero. This does not mean that these deities have lost connections, but that more connections elsewhere in the network have become available to navigate from one deity to another. Notably, Rovagug's betweenness decreases by more than half, hinting that there is more to these deities' relationships than merely uniting against the destruction of existence.

To compare whole-network structures, we turn to the node degree cosine, node strength cosine, and edge existence Jaccard. Comparing the LOGM and LODM networks results in $\textrm{NDC}^{\textrm{LOGM,LODM}} = 0.943$ and $\textrm{NSC}^{\textrm{LOGM,LODM}} = 0.952$, indicating a similar overall edge structure when scaled for network size. However, $\textrm{EEJ}^{\textrm{LOGM,LODM}} = 0.571$ indicates significant differences in the edges that are present in each network. Roughly speaking, the $43\%$ difference in edge existence between the two networks highlights the additional edges present in LODM.

\subsection{The Impacts of the Pathfinder Remaster, Gorum's Death, Arazni's Rise, and Increased Page Count}

As discussed in Section \ref{sec:intro}, four primary changes prompted differences between LOGM and LODM: the Pathfinder Remaster, the death of Gorum, Arazni's rise to prominence, and a doubling of the page count devoted to each deity. While it is impossible to fully disentangle the effects of these changes on the divine social network, we can get an idea of their relative impact by carrying out the following: First, we examine how the nodes cluster into communities to investigate whether deities of common alignment are more strongly associated with each other. Second, we consider subnetworks in which Gorum and Arazni are excluded.

\subsubsection{Alignment: Clustering Analysis}

Applying the Louvain method \cite{Louvain} to LOGM produces the following three clusters of deities. We count the number of alignment traits represented in each cluster to determine how well alignment accounts for inter-deity connections.
\begin{enumerate}
    \item Erastil, Gorum, Gozreh, Nethys, Pharasma, Urgathoa. This cluster includes 1 good deity, 1 evil deity, 1 lawful deity, and 1 chaotic deity. 
    \item Abadar, Calistria, Cayden Cailean, Iomedae, Irori, Norgorber, Torag. This cluster includes 3 good deities, 1 evil deity, 3 lawful deities, and 2 chaotic deities.
    \item Asmodeus, Desna, Lamashtu, Rovagug, Sarenrae, Shelyn, Zon-Kuthon. This cluster includes 3 good deities, 4 evil deities, 2 lawful deities, and 3 chaotic deities.
\end{enumerate}

This clustering seems appropriate given that it places the Ascended in the same cluster and the Radiant Prism in the same cluster. Overall, the clusters do not seem to partition the core 20 deities along alignments: Clusters 1 and 3 are mostly balanced between good and evil and between law and chaos. Cluster 2 features more good deities than evil deities, but this could be confounded with the fact that it includes the Ascended, which brings in 2 good deities and 1 evil. Cluster 1 includes some antagonistic relationships (Pharasma against Urgathoa, Gozreh against Nethys and Urgathoa) but this is not consistent as Asmodeus and Iomedae are in separate clusters. Therefore, it does not seem that alignment clearly explains deity affiliations in the social network.

\subsubsection{The Death of Gorum: LOGM and LOGM sans Gorum}

Comparing the densities of the networks from LOGM and LOGM sans Gorum, we find a slight increase in density with Gorum's removal: $\rho^{\textrm{LOGM}} = 0.337, \rho^{\textrm{LOGM sans Gorum}} = 0.357$. With Gorum removed from the network, there are fewer ``missing'' edges in the denominator of the density calculation.

We can examine the deities' centrality measures in Table \ref{tab:centralities} by comparing the LOGM columns and the LOGM sans Gorum columns. The deities' closeness values show little change, indicating that Gorum rarely provided a shorter path between two deities in LOGM. Similarly, the betweenness values are overall similar to each other, with two notable exceptions: Calistria's betweenness value increases by approximately $40\%$, and Norgorber's betweenness value reduces by more than half. It therefore seems that Gorum's removal from the pantheon makes Calistria more central to the network and Norgorber less central.

The overall network structure remains relatively undisturbed by Gorum's death. We find a node degree cosine of $\textrm{NDC}^{\textrm{LOGM,LOGM sans Gorum}} = 0.994$, a node strength cosine of $\textrm{NSC}^{\textrm{LOGM,LOGM sans Gorum}} = 0.996$, and an edge existence Jaccard of $\textrm{EEJ}^{\textrm{LOGM,LOGM sans Gorum}} = 0.953$. It therefore seems that, apart from shifts in betweenness for Calistria (an increase) and Norgorber (a decrease), Gorum's death had little impact on the divine social network in LOGM.

\subsubsection{The Rise of Arazni: LODM and LODM sans Arazni}

Comparing the densities of the networks from LODM sans Arazni and LODM, we find a slight increase in density with Arazni's addition: $\rho^{\textrm{LODM sans Arazni}} = 0.416, \rho^{\textrm{LODM}} = 0.421$. Even though Arazni claims to be a reluctant and disengaged goddess, she brings to the pantheon enough relationships to make the network slightly more connected.

We next examine the deities' centrality measures in Table \ref{tab:centralities} by comparing the LODM sans Arazni columns and the LODM columns. The deities' closeness values show little change, with Norgorber's closeness increasing by about $27\%$ thanks to Arazni's addition. Similarly, the betweenness values are overall similar to each other, although Iomedae's betweenness increases by a factor of 19 with the addition of Arazni. Cayden Cailena's betweenness increases by $80\%$, and Calistria's betweenness drops to zero.

We find a node degree cosine of $\textrm{NDC}^{\textrm{LODM,LODM sans Arazni}} = 0.981$ and a node strength cosine of $\textrm{NSC}^{\textrm{LODM,LODM sans Arazni}} = 0.972$, indicating little overall change. However, the edge existence Jaccard $\textrm{EEJ}^{\textrm{LODM,LODM sans Arazni}} = 0.911$ indicates a nearly $10\%$ change in edges thanks to Arazni's addition. 

\subsubsection{The Doubling of Page Counts: LOGM sans Gorum and LODM sans Arazni}

Finally, a comparison between LOGM sans Gorum and LODM sans Arazni helps to emphasize the impact of the additional page count devoted to each deity. It is between these two networks that we observe the most significant changes in betweenness centrality measures. Looking at the networks overall, we find a node degree cosine of $\textrm{NDC}^{\textrm{LOGM sans Gorum,LODM sans Arazni}} = 0.970$ and a node strength cosine of $\textrm{NSC}^{\textrm{LOGM sans Gorum,LODM sans Arazni}} = 0.981$. These values indicate more overall change than we observed when isolating Gorum's removal or Arazni's inclusion. The greatest difference lies in the edge existence Jaccard $\textrm{EEJ}^{\textrm{LOGM sans Gorum,LODM sans Arazni}} = 0.642$ indicates a sizable change in edges without Gorum or Arazni present. 

\section{\label{sec:discussion}Discussion}

Here we answer our research questions and discuss the notable differences between the networks reviewed in Section \ref{sec:results} and offer potential interpretations of these differences based on the setting and publishing changes that they accompanied. We then offer suggestions for future explorations in the Lost Omens campaign setting based on our observations, and finally reflect on limitations of our study.

\subsection{What Changed?}

In general, the social network resulting from LODM is richer than that of LOGM: There are more connections as reflected in the higher density values and general increase in deity centrality measures. There is some shifting in terms of which deities are most central, with Rovagug playing less of a role in connecting the deities that once allied against him, reflecting their own bona fide relationships apart from fighting the god of destruction.

We have attempted to disentangle the particular impacts of the removal of alignment, the death of Gorum, the rise of Arazni, and the increased page count between LOGM and LODM. We have made this attempt by examining clusters for evidence of alignment-based ties, creating a subnetwork of LOGM with Gorum removed, and creating a subnetwork of LODM with Arazni not yet introduced. We conceive of this as a chronological process, LOGM turning into LOGM sans Gorum, then becoming LODM sans Arazni, followed by the full LODM network. This process has revealed the following:

\textbf{We find no evidence of alignment-based clustering.} Overall, the clusters obtained using the Louvain algorithm do not strongly favor any particular alignment traits. One cluster contains more good deities than evil, but this association is confounded by the strong ties between the Ascended within that cluster. Therefore, even before the Pathfinder Remaster removed alignment from the game, deities do not seem to be bound to association simply because of alignment.

\textbf{Gorum's death shows limited impact on the network.} In the LOGM network, Gorum was already relatively unconnected from the pantheon. He had only 3 edges directly connecting him to other deities, and his centrality measures were relatively small. When removing him from the LOGM network, little changes in terms of the other deities' centrality measures (save Calistria's and Norgorber's betweenness values, which we discuss further below). More broadly, the NDC, NSC, and EEJ values between LOGM and LOGM sans Gorum confirm little change to the pantheon's social network structure due to Gorum's death. In retrospect, this makes Paizo's choice to kill off Gorum seem quite sensible, as it left them free to focus on the impact of Gorum's death within the mortal realm without needing to completely restructure the pantheon.

\textbf{Arazni's rise shows a slightly stronger impact on the network.} We see slightly greater differences between LODM sans Arazni and the full LODM network. Arazni's presence has a greater impact on other deities' centrality measures than did Gorum's departure. This could be due to the additional page count afforded Arazni in LODM than Gorum in LOGM, but we discuss the greater impacts of page counts next. Examining Arazni's strong connections with Iomedae and Shelyn---two deities already central to the network---she has entered the pantheon already more well-connected than Gorum was after millennia of godhood.

\textbf{The doubling of page counts strongly impacted the network.} The impacts of Gorum's removal and Arazni's rise are dwarfed by the impact of the doubling of page count between LOGM and LODM, reflected in the differences between LOGM sans Gorum and LODM sans Arazni. We see the greatest shifts in centrality measures between these two subnetworks, as well as the greatest difference in edge existence ($\textrm{EEJ} = 0.642$). It seems that with the additional page counts afforded in LODM, Paizo chose to develop more of the inter-deity relationships by nearly double.

\textit{The next paragraph contains spoilers for events surrounding the War of Immortals and the Curtain Call adventure path.} 

\textbf{We find a curious case with Calistria and Norgorber.} While most centrality measures remained undisturbed by Gorum's death, we find it curious that the largest changes would befall the betwenneess centrality measures of Calistria and Norgorber. Calistria, who arranged Gorum's death at his request, rose in betweenness as a result of Gorum's death. Meanwhile Norgorber, the god of secrets who used Gorum's death in a bid to grow in power, became less centralized in the divine network, perhaps reinforcing his secretive positioning. Interestingly, the introduction of Arazni to the pantheon has the opposite effect: Between LODM sans Arazni and LODM, Calistria's betweenness value decreases to zero and Norgorber's closeness value increases. Presuming Calistia and Norgorber preferred the changes to the network brought about by the death of Gorum, Arazni could very well find herself at odds with Calistria and Norgorber for thwarting those changes.

\subsection{Paths Still to be Found}

The LODM network in Figure \ref{fig:network-diagrams}(b) depicts the current state of divine social network connections, but also draws attention to connections yet to be explored in Lost Omens lore.

\textbf{Nethys is relatively disconnected.} As the god of magic, Nethys holds a prominent place philosophically in the pantheon, yet is the most disconnected of the Lost Omens deities. With a small closeness and zero betweenness, Nethys' relationship with the other deities is waiting to be explored. 

\textbf{Shelyn is highly centralized.} In the LODM network, Shelyn enjoys the highest closeness and betweenness values of all the deities. This is largely due to several strong edges that connect her to Sarenrae and Desna (the Radiant Prism), Zon-Kuthon (her brother), newcomer Arazni (whom Shelyn supports with care), and Urgathoa (philosophical rival). These ties could set up Shelyn to be the center of a major event in the setting.

\textbf{There is potential for holy alliances and unholy rivalries.} Neither the LOGM nor the LODM network heavily favors relationships between deities based on alignment or sanctification. For example, Iomedae, Torag, and Erastil all previously featured the same alignment (lawful good) and now align in sanctification options (either requiring holy sanctification or offering only holy sanctification). However, neither network shows a strong edge between their nodes, leaving the nature of their alliances or common interest in the fight against evil underexplored. Similarly, deities requiring or offering only unholy sanctification (Asmodeus, Lamasthu, Norgorber, Urgathoa, and Zon-Kuthon) are connected with weaker edges. Some of these edges represent rivalries, which could be further explored, perhaps in adventures in which the player characters play these deities' followers against each other.

\subsection{Limitations}

The primary limitation in this study lies in the data collection. Even with expanded page count in LODM, space requirements limit how many divine connections can be established in Pathfinder game materials. There are certainly more inter-deity interactions specified throughout Lost Omens lore and summarized in the Pathfinder Wiki, but we have chosen to focus on LOGM and LODM because of their similar purpose and their consistent structure across the deities' profiles.

To collect our data, we have used an electronic word search feature to find and count the number of occurrences of deity names in each deity's profile. In so doing, we have used each deity's proper name at the time of present day in the setting, omitting any references to deity titles (e.g., ``the Rough Beast'' instead of ``Rovagug'') or to former names (e.g., ``Dou-Bral'' instead of ``Zon-Kuthon'').

Finally, our analysis has attempted to disentangle the effects of page count increase and the adjustment of pantheon members. In doing so, we constructed an artificial timeline by removing Gorum and Arazni from the LOGM and LODM networks, respectively. While we found that Arazni's introduction had a greater impact than Gorum's removal, we also must acknowledge that Arazni received twice as many pages for her profile as Gorum. This approach also does not account for Paizo's holistic decision-making process behind removing Gorum, incorporating Arazni, and increasing the page count for deity profiles.

\section{Conclusions}

We have used social network analysis to examine the differences between inter-deity connections in the Lost Omens campaign setting between \textit{Pathfinder: Lost Omens Gods \& Magic} and \textit{Pathfinder: Lost Omens Divine Mysteries}. These changes are brought about by the Pathfinder Remaster, the death of Gorum, the rise of Arazni to the setting's primary pantheon, and the doubling of page count devoted to each deity's profile. We have attempted to disentangle these effects through clustering analysis and by creating subnetworks without Gorum and Arazni. By examining various network comparison metrics, we find that the doubling of page count has likely had the most significant impact on the network, followed by a smaller impact by the introduction of Arazni, the smallest impact by Gorum's death, and no impact by the Pathfinder Remaster's removal of alignment. We used this analysis to offer suggestions for future development in the divine social network.

% \begin{figure}
%     \centering
%     \includegraphics[width=1\linewidth]{networks-portrait.png}
%     \caption{Enter Caption}
%     \label{fig:enter-label}
% \end{figure}

\acknowledgements{This work was carried out on Google Colab, with a cloneable Jupyter notebook available at reference \cite{Jupyter}. We are grateful to Paizo for their support of the Infinite Masters program. Images are used under the Paizo Community Use Policy.}

\bibliography{apssamp}

\providecommand{\noopsort}[1]{}\providecommand{\singleletter}[1]{#1}%
\begin{thebibliography}{6}%
\makeatletter
\providecommand \@ifxundefined [1]{%
 \@ifx{#1\undefined}
}%
\providecommand \@ifnum [1]{%
 \ifnum #1\expandafter \@firstoftwo
 \else \expandafter \@secondoftwo
 \fi
}%
\providecommand \@ifx [1]{%
 \ifx #1\expandafter \@firstoftwo
 \else \expandafter \@secondoftwo
 \fi
}%
\providecommand \natexlab [1]{#1}%
\providecommand \enquote  [1]{``#1''}%
\providecommand \bibnamefont  [1]{#1}%
\providecommand \bibfnamefont [1]{#1}%
\providecommand \citenamefont [1]{#1}%
\providecommand \href@noop [0]{\@secondoftwo}%
\providecommand \href [0]{\begingroup \@sanitize@url \@href}%
\providecommand \@href[1]{\@@startlink{#1}\@@href}%
\providecommand \@@href[1]{\endgroup#1\@@endlink}%
\providecommand \@sanitize@url [0]{\catcode `\\12\catcode `\$12\catcode `\&12\catcode `\#12\catcode `\^12\catcode `\_12\catcode `\%12\relax}%
\providecommand \@@startlink[1]{}%
\providecommand \@@endlink[0]{}%
\providecommand \url  [0]{\begingroup\@sanitize@url \@url }%
\providecommand \@url [1]{\endgroup\@href {#1}{\urlprefix }}%
\providecommand \urlprefix  [0]{URL }%
\providecommand \Eprint [0]{\href }%
\providecommand \doibase [0]{https://doi.org/}%
\providecommand \selectlanguage [0]{\@gobble}%
\providecommand \bibinfo  [0]{\@secondoftwo}%
\providecommand \bibfield  [0]{\@secondoftwo}%
\providecommand \translation [1]{[#1]}%
\providecommand \BibitemOpen [0]{}%
\providecommand \bibitemStop [0]{}%
\providecommand \bibitemNoStop [0]{.\EOS\space}%
\providecommand \EOS [0]{\spacefactor3000\relax}%
\providecommand \BibitemShut  [1]{\csname bibitem#1\endcsname}%
\let\auto@bib@innerbib\@empty
%</preamble>
\bibitem [{\citenamefont {Hoffer}(2022)}]{Hoffer2022Dungeons}%
  \BibitemOpen
  \bibfield  {author} {\bibinfo {author} {\bibfnamefont {C.}~\bibnamefont {Hoffer}},\ }\bibfield  {title} {\bibinfo {title} {Dungeons \& dragons announces changes to ogl, some third-party creators must report revenue and potentially pay royalties},\ }\href {https://comicbook.com/gaming/news/dungeons-dragons-wizards-of-the-coast-ogl-srd-royalties/} {\bibfield  {journal} {\bibinfo  {journal} {ComicBook.com}\ } (\bibinfo {year} {2022})}\BibitemShut {NoStop}%
\bibitem [{\citenamefont {Codega}(2023)}]{Codega2023Paizo}%
  \BibitemOpen
  \bibfield  {author} {\bibinfo {author} {\bibfnamefont {L.}~\bibnamefont {Codega}},\ }\bibfield  {title} {\bibinfo {title} {Paizo announces a new gaming license amid dungeons \& dragons' ogl controversy},\ }\href {https://gizmodo.com/paizo-wizards-of-the-coast-dnd-open-rpg-ogl-1-1-1849982443} {\bibfield  {journal} {\bibinfo  {journal} {Gizmodo}\ } (\bibinfo {year} {2023})}\BibitemShut {NoStop}%
\bibitem [{\citenamefont {Ferron}\ and\ \citenamefont {Loza}(2024)}]{LODM}%
  \BibitemOpen
  \bibfield  {author} {\bibinfo {author} {\bibfnamefont {E.}~\bibnamefont {Ferron}}\ and\ \bibinfo {author} {\bibfnamefont {L.}~\bibnamefont {Loza}},\ }\href@noop {} {\emph {\bibinfo {title} {Pathfinder: Lost Omens Divine Mysteries}}}\ (\bibinfo  {publisher} {Paizo},\ \bibinfo {year} {2024})\BibitemShut {NoStop}%
\bibitem [{\citenamefont {Ferron}\ and\ \citenamefont {Loza}(2020)}]{LOGM}%
  \BibitemOpen
  \bibfield  {author} {\bibinfo {author} {\bibfnamefont {E.}~\bibnamefont {Ferron}}\ and\ \bibinfo {author} {\bibfnamefont {L.}~\bibnamefont {Loza}},\ }\href@noop {} {\emph {\bibinfo {title} {Pathfinder: Lost Omens Gods \& Magic}}}\ (\bibinfo  {publisher} {Paizo},\ \bibinfo {year} {2020})\BibitemShut {NoStop}%
\bibitem [{Lou()}]{Louvain}%
  \BibitemOpen
  \href@noop {} {\bibinfo {title} {Louvain method}},\ \bibinfo {note} {\url{https://en.wikipedia.org/wiki/Louvain_method}}\BibitemShut {NoStop}%
\bibitem [{\citenamefont {Lane}()}]{Jupyter}%
  \BibitemOpen
  \bibfield  {author} {\bibinfo {author} {\bibfnamefont {W.~B.}\ \bibnamefont {Lane}},\ }\href@noop {} {\bibinfo {title} {Pf deities network.ipynb}},\ \bibinfo {note} {\url{https://colab.research.google.com/drive/1es5jiEs2tmbjYLkWhY4q\_mzCXsZaHgBt?usp=sharing}}\BibitemShut {NoStop}%
\end{thebibliography}%

\end{document}